\documentclass[10pt,english,aps,manuscript,prb,floatfix,superscriptaddress,twocolumn,showpacs,amsmath,amssymb,reprint]{revtex4-1}
\usepackage[T1]{fontenc}
\usepackage[latin9]{inputenc}
\usepackage{color}
\usepackage{amssymb}
\usepackage{graphicx}
\usepackage{esint}
\PassOptionsToPackage{normalem}{ulem}
\usepackage{ulem}

\makeatletter

\newcommand{\lyxdot}{.}

\@ifundefined{textcolor}{}
{%
 \definecolor{BLACK}{gray}{0}
 \definecolor{WHITE}{gray}{1}
 \definecolor{RED}{rgb}{1,0,0}
 \definecolor{GREEN}{rgb}{0,1,0}
 \definecolor{BLUE}{rgb}{0,0,1}
 \definecolor{CYAN}{cmyk}{1,0,0,0}
 \definecolor{MAGENTA}{cmyk}{0,1,0,0}
 \definecolor{YELLOW}{cmyk}{0,0,1,0}
}

\PassOptionsToPackage{caption=false}{subfig}

\@ifundefined{showcaptionsetup}{}{%
 \PassOptionsToPackage{caption=false}{subfig}}
\usepackage{subfig}
\makeatother

\usepackage{babel}
\begin{document}

\title{Majorana Fermions Signatures in Macroscopic Quantum Tunneling}
\begin{abstract}
Thermodynamic measurements of the magnetic flux and I-V characteristics
in SQUIDs offer promising paths to the characterization of topological
superconducting phases.Through a simplified model, we consider the effects
of topological superconducting phases on the macroscopic quantum
behavior of an rf-SQUID. We show that the topological order shifts 
the tunneling rates and quantum levels, both in the parity conserving 
and fluctuating cases. The latter case is argued to actually enhance 
the signatures in the slowly fluctuating limit, which is expected to take 
place in the quantum regime of the circuit. In view of recent advances, 
we also discuss how our results affect a $\pi$-junction loop.
\end{abstract}

\author{Pedro L. e S. Lopes}

\email{plslopes@ifi.unicamp.br}

\affiliation{Instituto de F\'{i}sica Gleb Wataghin, Universidade Estadual de Campinas,
Campinas, SP 13083-970, Brazil}

\affiliation{Department of Physics and Institute for Condensed Matter Theory,
University of Illinois, 1110 W. Green St., Urbana IL 61801-3080, U.S.A.}

\author{Vasudha Shivamoggi }

\altaffiliation[Present location: ]{Northrop Grumman Electronic Systems, Linthicum Heights, MD 21090}
\affiliation{Department of Physics and Institute for Condensed Matter Theory,
University of Illinois, 1110 W. Green St., Urbana IL 61801-3080, U.S.A.}

\author{Amir O. Caldeira}

\affiliation{Instituto de F\'{i}sica Gleb Wataghin, Universidade Estadual de Campinas,
Campinas, SP 13083-970, Brazil}


\maketitle

\section{Introduction}

Global symmetries have important consequences in the characterization
of phases of matter. Topological superconductors (TSC), for example,
are systems in which a global particle-hole symmetry protects robust
edge states which are predicted to be Majorana fermions, particles
known for being their own anti-particles. These predictions, however,
still lack some experimental evidence and, despite the recent  efforts
of the condensed matter community \cite{Beenakker2013a,Alicea2012},
no conclusion has been reached so far.

Majorana bound states (MBS) are immune to electromagnetic influences
and this, along with their braiding properties and the possibility
of forming non-local complex fermions, make them perfect candidates
to be used in quantum computation platforms\cite{MajoQComp}. The
immunity to electromagnetic probing and the fragility of the TSC phase,
however, make the experimental unveiling of these particles rather
difficult.

Typical approaches in the search for evidence of Majorana fermions
involve transport experiments and the probing of zero-bias peaks.
These are, on the other hand, typically plagued by ambiguities in
the interpretation of the results. A complementary approach to these,
 based in thermodynamical measurements, is desirable, avoiding
the aforementioned ambiguities.

Macroscopic quantum phenomena may provide such an alternative process.
In particular, mesoscopic rf-SQUIDs have been shown to possess a quantum
regime \cite{Friedman2000,Makhlin2001} in which the flux through
the SQUID ring (generated by a macroscopic current) fluctuates quantum
mechanically. The reading of this flux is exactly such a thermodynamic
type of measurement which avoids transport phenomena. It is our main
goal in this work to describe a scenario in which MBS physics and
macroscopic quantum phenomena are connected. We discuss imprints of
MBS in the macroscopic quantum tunneling (MQT) behavior of the magnetic
flux in a TSC loop in its quantum regime. 

Besides the difficulties with experimental signatures, the very realization
of TSC is, by itself, a challenge. Up to date, no superconducting
(SC) material is known to develop naturally its topological regime.
Analogous phases which behave, in all aspects, as TSCs have been proposed
and realized like the 5/2 state in fractional quantum Hall effect\cite{52HallTheory,Willett02062009}.
Strategies proposed for realizing TSC involve the use of proximity effects between trivial s-wave SCs
and other strong spin-orbit coupled materials. Promising approaches
are the coupling of s-wave SCs to the helical modes along the edges of quantum
spin-Hall insulators (QSHI) \cite{Fu2009,Beenakker2013} and the coupling
of s-wave SCs to quasi-1D nanowires of semiconductors with strong
spin-orbit (SO) coupling \cite{PhysRevLett.105.077001,PhysRevLett.105.177002,Alicea2011}

\begin{figure}[t]
\centering{}\subfloat[\label{fig:SQUID}]{\begin{centering}
\includegraphics[scale=0.33]{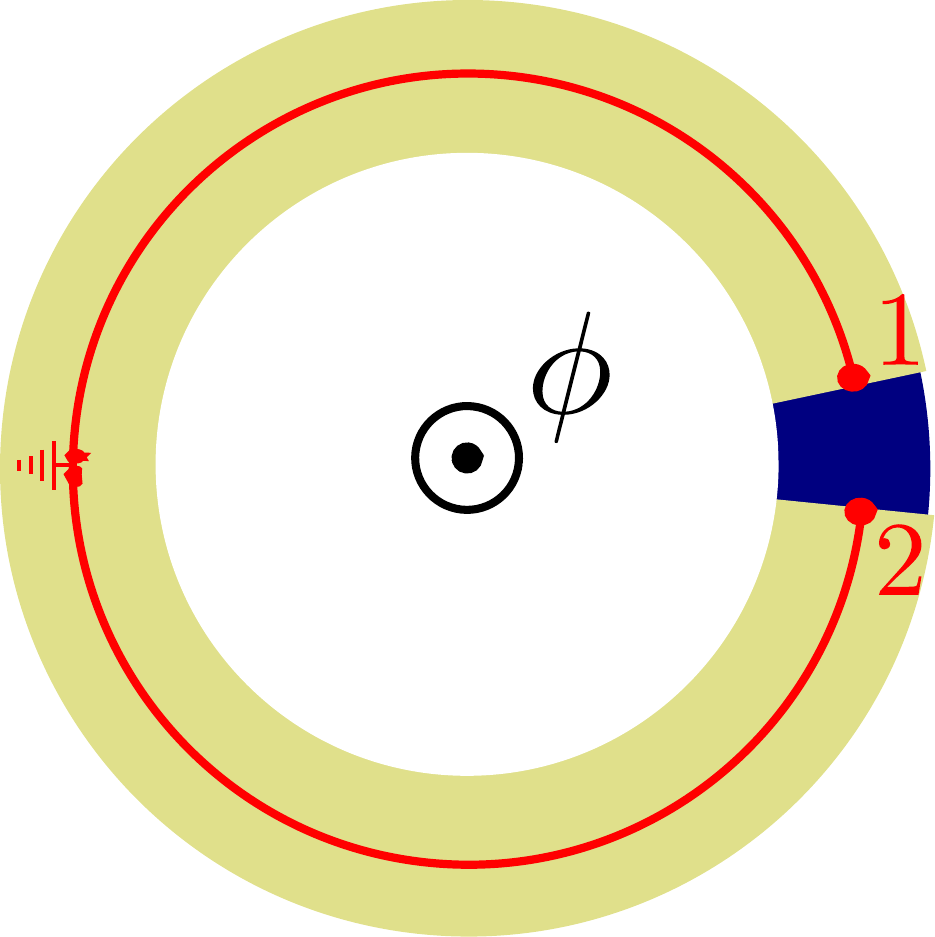}
\par\end{centering}

}\subfloat[\label{fig:LumpedSQUID}]{\begin{centering}
\includegraphics[scale=0.22]{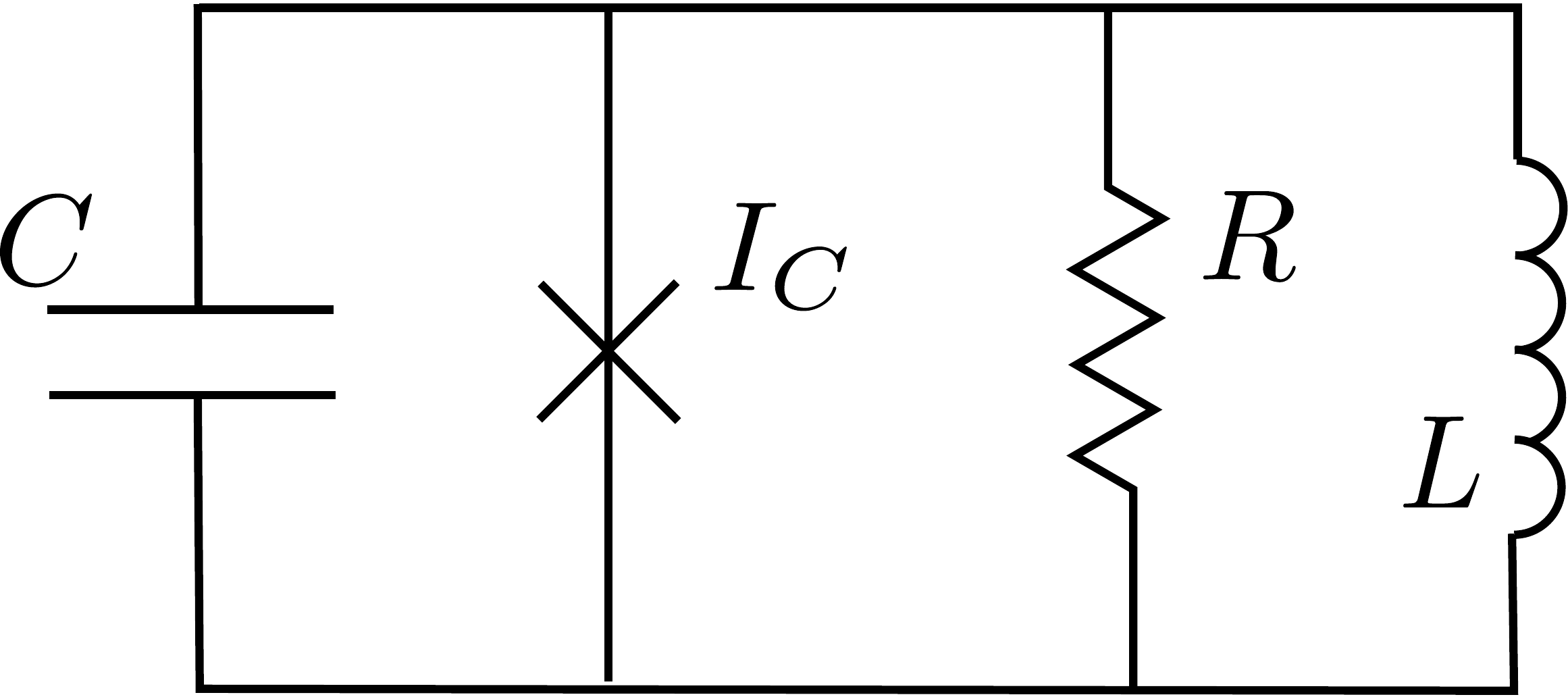}
\par\end{centering}

}\protect\caption{(Color online) (a) rf-SQUID with topological component device schematics.
The beige ring represents a ''parent'' s-wave SC, and the blue region
is an insulating barrier that creates a Josephson junction.The red
ring represents a semiconducting wire with strong spin-orbit coupling
that is proximity-coupled to the parent superconductor in all its length. We represent
 the possibility of exchange of quasi-particles between the rings by the grounding symbol; 
dots represent Majorana fermions at the junction;
(b) Lumped circuit representation of an rf-SQUID (no-topological component
here). The circuit consists of a capacitive component connected in
parallel with a resistor, an inductor and a JJ.}
\end{figure}

We depict in Fig.\ref{fig:SQUID} the simplified model on which we focus in
this work. Imagine a Josephson junction (JJ) consisting of a strongly
spin-orbit (SO) interacting wire (red) lying over a s-wave SC broken
ring (beige). We call the latter the ''parent'' SC, responsible
to induce p-wave pairing and TSC in the SO wire. 

In the presence of a magnetic field, the wire effectively develops
topological superconductivity and may be described as a Kitaev chain
\cite{KitaevChain}. MBSs arise at the edges of a Kitaev wire and
induce the so-called $4\pi$ periodic topological Josephson effect
when two wires are allowed to couple \cite{Alicea2012,PhysRevB.84.180502,Fu2009,KitaevChain}.
In our setup, we imagine the parent SC touching the wire along the whole of its length and
acting as grounding for the chain. In such an example, we expect the $4\pi$ periodicity to develop in the
wire's JJ, even if it is a single one. Our particular choice of device,
nevertheless, is not fundamental and, as we will discuss later on in this manuscript, similar physics
would arise in other situations, as in the SC-QSHI-SC junction.

The $4\pi$ periodicity has striking consequences in macroscopic quantum
phenomena. Majorana particles mediate tunneling only between even
quantum flux states. This picture is to be contrasted with the trivial
JJ situation which allows for tunneling between states of any integer
number of flux quanta. This means that the tunneling barrier in the topological
case is much wider than that for the non-topological one. Probing
the topological phase transition then may be done comparing the changes
in the tunneling rate of the device. In our device, however, the coupling
between the topological wire and the parent SC introduces a dominant
$2\pi$ ``trivial'' Josephson energy to the $4\pi$ periodic one
and we must study the interplay between these.

A possible issue concerning our signatures is that the pair of MBSs
at the junction define a two-level system characterized by its occupancy
through a fermionic parity observable. In real systems this parity
conservation is frequently broken. Defects and leads are sources for
stray quasi-particles that may couple to the edge modes and change
their fermionic parity state. These phenomena, generally dubbed quasi-particle
poisoning, have been discussed in the past and are usually blamed
for being responsible for washing away the $4\pi$ periodic signatures.
These effects, however, have been shown to induce other signatures
like telegraph noise\cite{Fu2009} and multiple critical currents\cite{Shu-Ping:1403.2747v1}
in open wire geometries. 

As a first approach, we focus ourselves on the treatment of the parity
conserving limit pointing out how MQT and spectroscopy experiments
may uncover the TSC phase. It is impossible, however to leave the
quasi-particle poisoning issue without any comment. The characteristic
time scales of fluctuations and tunneling have to be considered carefully.
We adopt then an heuristic point of view and address what are the
expected effects of parity fluctuations in the MQT signatures. We
argue that the latter are expected to be quite robust against the
quasi-particle poisoning and may even be enhanced, as long as the
fluctuations are slower than the tunneling processes.

The paper is organized as follows. We start in Section \ref{sec:review}
with a short review of how the subject arises in
the context of flux dynamics in rf-SQUIDs. In Section \ref{sec:opendwire}
we introduce, justify and thoroughly explain the phenomenological
model. In Section\ref{sec:signatures} we describe the predicted signatures
in tunneling and resonance experiments and show our main results.
We discuss the effects of parity fluctuations and quasi-particle poisoning
in Section \ref{sec:Parityfluctuations}. We save Section \ref{sec:pi-junction}
to address briefly the $\pi$-junction limit and show how the main
results of the previous sections would change. We close in Section
\ref{sec:Conclusions} with our conclusions.

\section{rf-SQUID and MQT\label{sec:review}}

\begin{figure}
\begin{centering}
\includegraphics[scale=0.5]{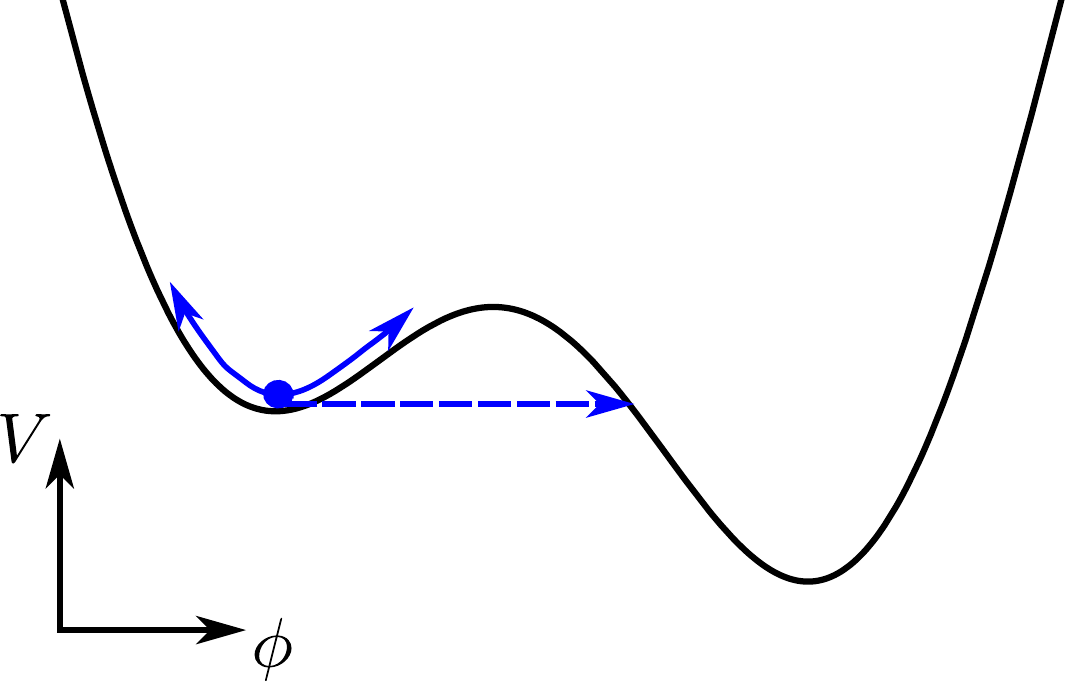}
\par\end{centering}

\protect\caption{Potential energy (\ref{eq:Upot}) for a particle with position ``$\phi$'',
actually the magnetic flux, according to the rf-SQUID equations of
motion.\label{fig:PotSchem}}
\end{figure}

We start with a brief review of the rf-SQUID and MQT phenomenology,
which may be skipped by readers familiar with the subject.

An rf-SQUID consists of a SC loop ring interrupted by a narrower region
or insulating barrier, which gives rise to a JJ. This is depicted
in Fig.\ref{fig:SQUID} as the beige ring. The physics of the whole
device is very successfully modeled by an RLC circuit with a JJ circuit
element \cite{LikharevBook,TinkhamBook} as in Fig.\ref{fig:LumpedSQUID}.
This is known as the resistor-capacitor shunted junction (RCSJ) model.

This model describes the interplay of the capacitive (kinetic), resistive
(from leads and normal current components present around the loop,)
(self-)inductive and JJ current contributions to the flux piercing
the ring. Current conservation through the circuit and Faraday's law
results in the equation of motion
\begin{equation}
C\ddot{\Phi}+\frac{\dot{\Phi}}{R}+I_{C}\sin\Delta\theta=\frac{\Phi_{X}-\Phi}{L}+\zeta\left(t\right),
\end{equation}
where $C$ is the capacitance of the junction, $R$ is its resistance
in the normal state, $I_{C}$ is the junction critical current, and
$\zeta\left(t\right)$ is a fluctuating current represented by a delta
correlated thermal noise. We have also considered the possibility
of adding an externally controlled flux $\Phi_{X}$ through the ring.
The phase difference $\Delta\theta$ across the junction may be related
to the magnetic flux in the closed geometry by the usual flux quantization
rule. For a broken SC loop it gives\cite{AmirBook}
\begin{equation}
\Phi+\frac{\Phi_{0}}{2\pi}\Delta\theta=n\Phi_{0},\label{eq:fluxquant}
\end{equation}
as long as the SC is thicker than the London penetration depth.

In this way, the equation of motion reduces to
\begin{equation}
C\ddot{\Phi}+\frac{1}{R}\dot{\Phi}+U^{'}\left(\Phi\right)=\zeta\left(t\right),
\end{equation}
which is a Langevin equation of motion for a classical dissipative
particle (with coordinate $\Phi$) in a conservative potential $U\left(\Phi\right)$
given by
\begin{equation}
U\left(\Phi\right)=U_{0}\left[\frac{\left(2\pi\left(\Phi-\Phi_{X}\right)\right)^{2}}{2}-\beta_{L}\cos\left(2\pi\Phi\right)\right],\label{eq:Upot}
\end{equation}
where $U_{0}=\frac{\phi_{0}^{2}}{4\pi^{2}L}$, $\beta_{L}=\frac{2\pi Li_{0}}{\phi_{0}}$
and with $\Phi$ (here and henceforth) measured in units of $\Phi_{0}=h/2e$.

As a first approach, we neglect dissipation (and noise) and focus
on the conservative part of the system in this work. The potential
is depicted in Fig.\ref{fig:PotSchem} for some arbitrary values of
$\beta_{L}$ and $\Phi_{X}$. As long as the Josephson energy (i.e.
$\beta_{L}$) is comparable to the inductive energy, ripples develop
in the parabolic potential, giving rise to local metastable minima.

For high enough temperatures, the flux may be thermally excited and
will slip to lower minima. Each minimum defines an oscillation frequency
\begin{equation}
\omega_{0}=\sqrt{\frac{1}{C\Phi_{0}^{2}}\frac{\partial^{2}U}{\partial\Phi^{2}}|_{\phi=\phi_{min}}}\sim\sqrt{\frac{1}{LC}},\label{eq:Freq}
\end{equation}
from which a characteristic temperature may be defined as
\begin{equation}
T_{0}\equiv\frac{\hbar\omega_{0}}{k_{B}}=0.76\times10^{-11}s\sqrt{\frac{1}{LC}}K.
\end{equation}
Parameters like $C\sim10^{-12}$ F, $L\sim10^{-10}$ H lead to $T_{0}\sim1$ K
(these parameters also lock $I_{C}$ to $\sim10^{-5}$ A). This means
that if the system is set at temperatures lower than $T_{0}$, it
may resolve the discrete energy levels within the metastable wells.
In this case, even if temperatures are much lower than the barrier
height, the flux may still escape to lower energy wells, now due to
quantum tunneling. This is the macroscopic quantum tunneling phenomenon.
\cite{1983AnPhy.149..374C,LeggettQMTest}
\begin{figure}[t]
\includegraphics[scale=0.24]{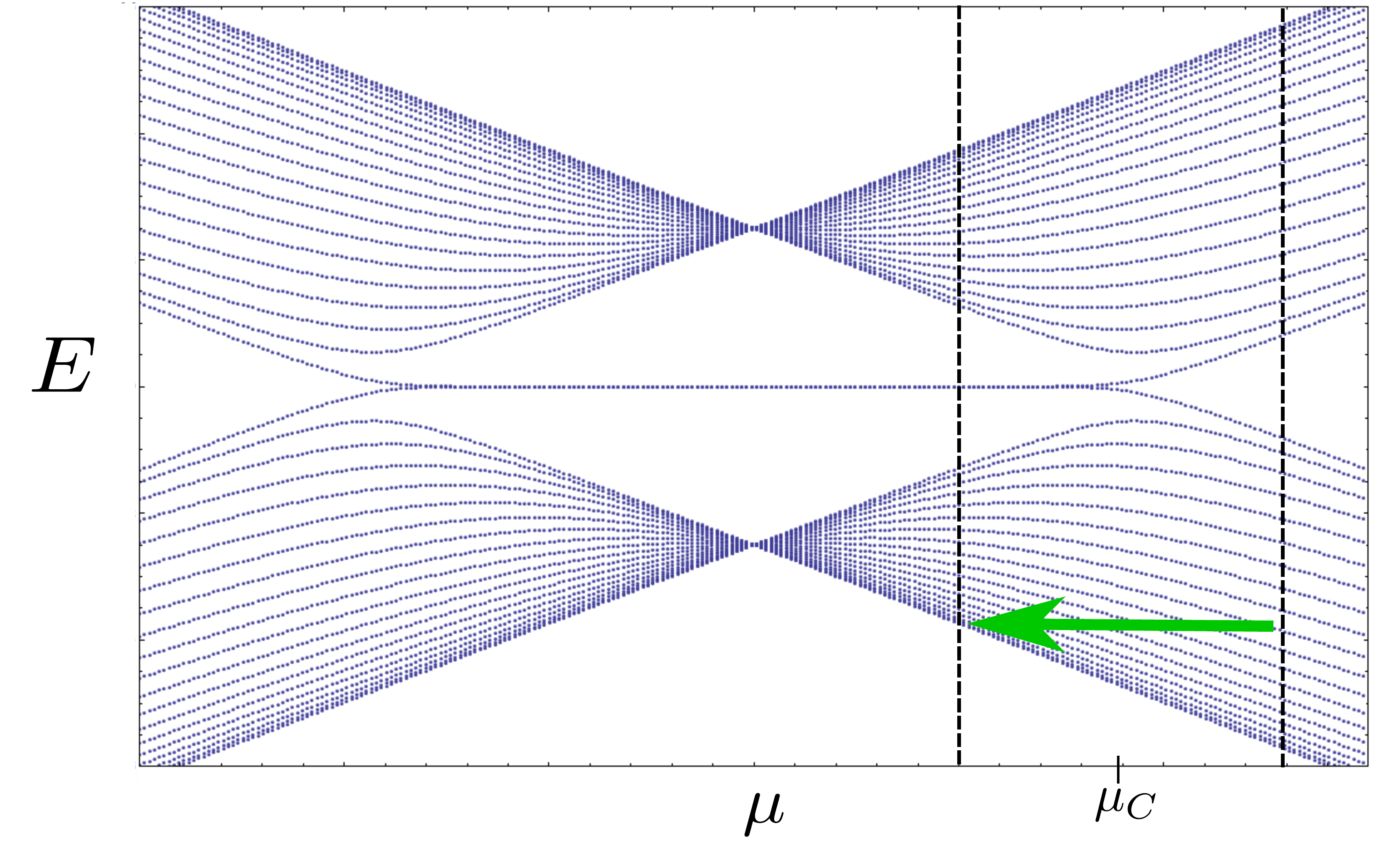}

\includegraphics[scale=0.28]{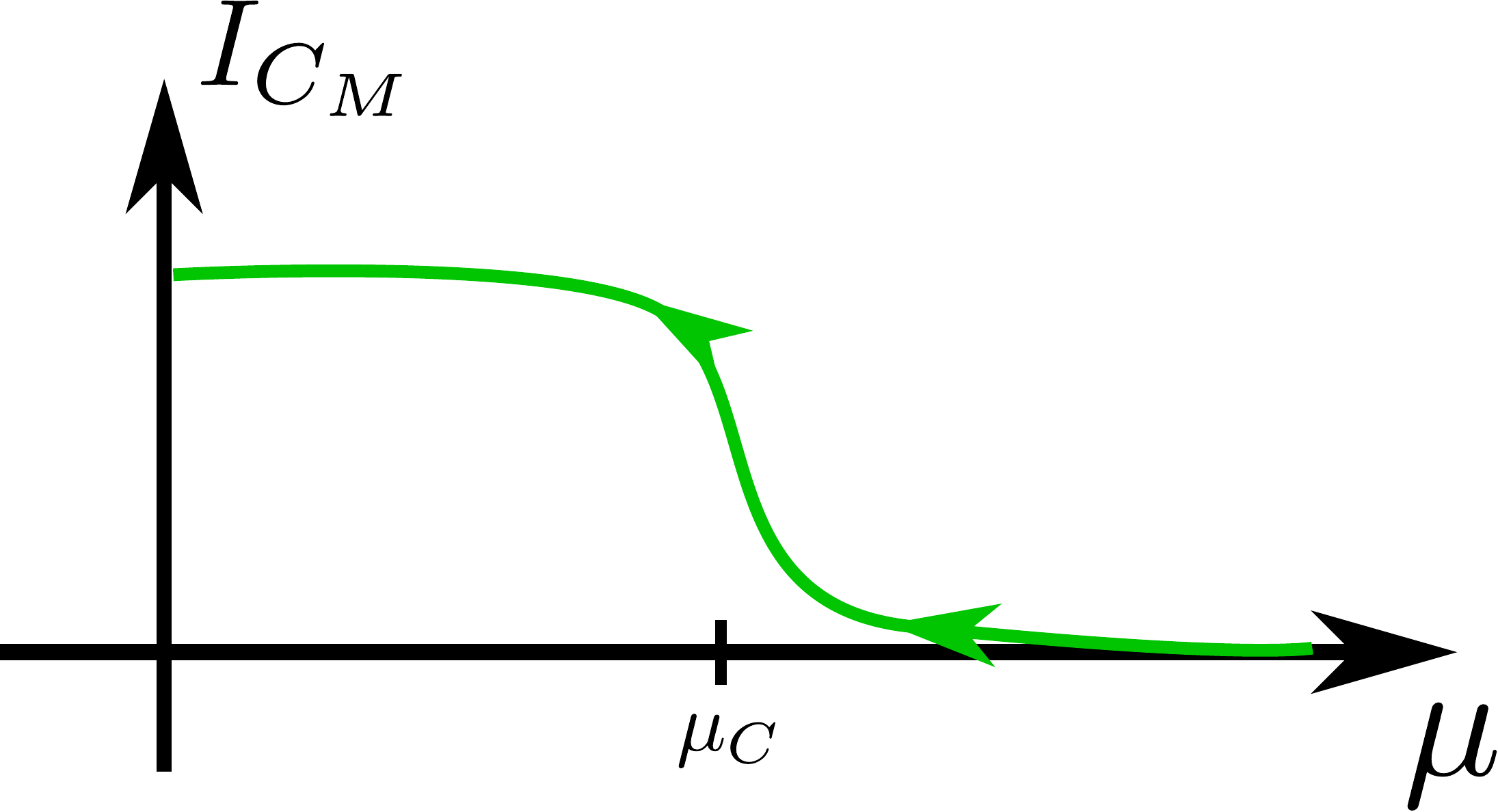}

\protect\caption{Schematic description of the behavior of the topological
 contribution to the SQUID critical current. 
(top) Schematic energy spectrum of an open Kitaev chain at
unit hopping and superconducting pairing. Dashed lines cut the spectrum pointing that
for low chemical potential the topological superconducting phase develops, as indicated by the
midgap flat-band; (bottom) Topological contribution to the critical
current as function of the chemical potential; the green arrow represents the tuning of
the chemical potential into the topological regime.\label{fig:CritCur}}
\end{figure}

\section{MBS signatures in MQT\label{sec:opendwire}}

To model the coupled topological and trivial SQUIDs, we start assuming
fermionic parity conservation. The wire in a topological phase allows
for the introduction of a topological JJ term in addition to the usual
Josephson current. The wire is assumed to be much thinner than the
SC ring, so that the capacitance and inductance of the device are
predominantly defined by the corresponding values from the parent
SC and do not depend much on the chemical potential of the wire. We
also assume the linear dimensions of the parent SC to be longer than
the SC penetration depth in such a way that the SQUID flux ``quantization''
condition is not changed. Finite-size effects of the ring are not
taken into account.

The coupling between the wire and parent SC serves as grounding for
the wire and allows for the parity anomaly. The full potential energy
thus has $2\pi$ and $4\pi$ periodic contributions which
compete for making $\Phi$ (close to) an arbitrary or even-only integer.
We consider then a new JJ element to the RCSJ model. The conditions
of thin wire guarantee that the phase across the topological JJ is
also controlled by the phase across the parent SC junction. The new
potential energy of the problem becomes 
\begin{eqnarray}
U\left(\phi\right) & = & U_{0}\left[\frac{\left(2\pi\left(\Phi-\Phi_{X}\right)\right)^{2}}{2}\right.\label{eq:pot}\\
 &  & \left.-\beta_{L}\left(\cos2\pi\Phi+\eta\left(\mu\right)\cos\pi\Phi\right)\right].\nonumber 
\end{eqnarray}
Here, $\eta\left(\mu\right)$ is a parameter given by the ratio $I_{C_{M}}/I_{C}$
between the critical currents of the parent s-wave junction and the
topological one. Its magnitude is roughly controlled by the ratio
between the magnitudes of the parent SC gap and the induced p-wave
gap in the wire. It will depend on the strength of the proximity effect
and on the parent SC 2D density of states. For high chemical potentials,
the 1D wire is in a trivial SC phase, whereas for low chemical potentials
it enters the topological regime \cite{Alicea2011}.

A subtlety concerns the sign of $\eta$. It is determined by which
parity sector the system is in \cite{Alicea2011}, and, as we assumed
the fermionic parity to be conserved, is fixed to a given value along
a complete tunneling process.We will come back to this point and address
the possibility of fluctuations of this occupancy of the non-local
two-level system generated by the MBSs.

Taking all that in consideration, we treat $\eta\left(\mu\right)$
phenomenologically. Assuming that, through gating, we may tune the
chemical potential, $\eta$ changes from zero to a saturated value
as the chemical potential moves from the trivial to the topological
regime. This general behavior is depicted in Fig.\ref{fig:CritCur}.
As the chemical potential goes from smaller to larger values, the
Majorana edge modes penetrate the bulk of the wire and, when in the
trivial phase, end up coupling and generating a complex fermion which
annihilates the topological contribution. In the trivial regime the
wire may give a small contribution to the $2\pi$ Josephson energy.
We neglect these effects assuming that whatever $2\pi$ periodic contribution
there may be, it is already included in $\beta_{L}$.

This potential also assumes a short junction. In the long junction
limit, more bound states develop at the junction and the physics becomes
more complicated \cite{Fu2009,Beenakker2013}.

Since the wire is much thinner than the s-wave SC and since the p-wave
pairing induced in the wire depends on the proximity coupling, it
is reasonable to assume that the saturated value of $\eta$ is not
very large and we will focus our quantitative discussions on this
case. On the other hand, by adding to the the JJ of the parent SC
a secondary loop, we may actually control the value of $\beta_{L}$
\cite{Friedman2000}, and as such, of $\eta$, thus allowing for some
control on this parameter.

\begin{figure}[t]
\begin{centering}
\includegraphics[scale=0.7]{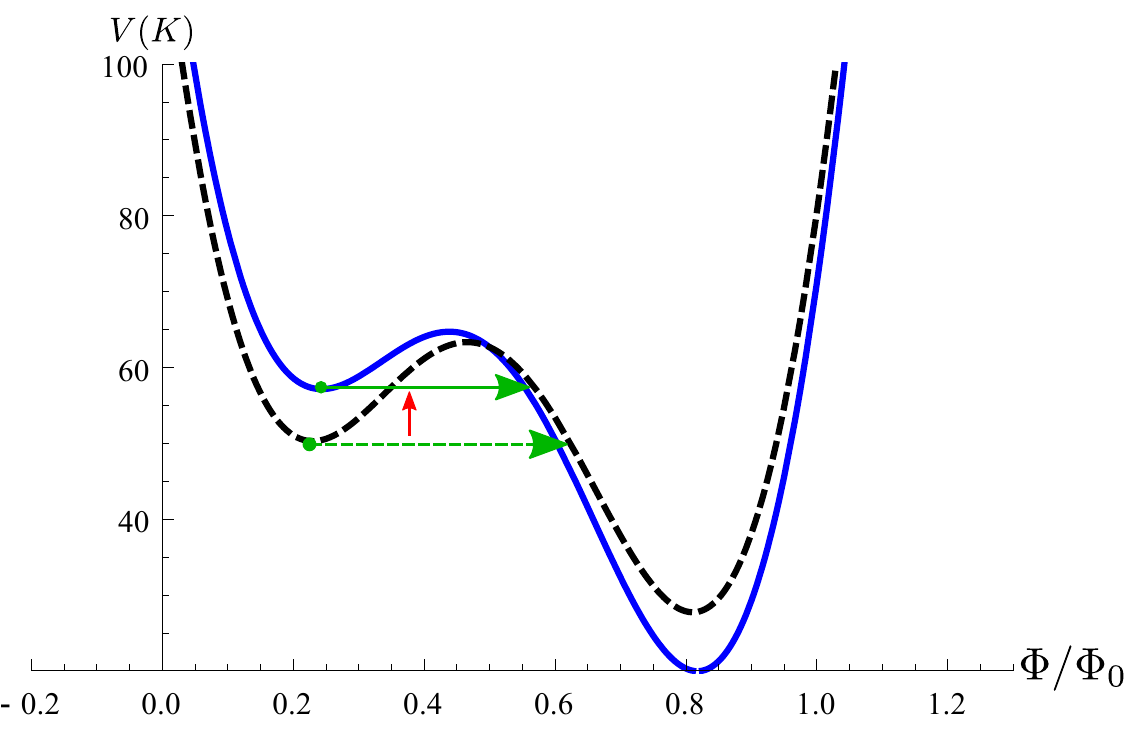}
\par\end{centering}

\begin{centering}
\includegraphics[scale=0.7]{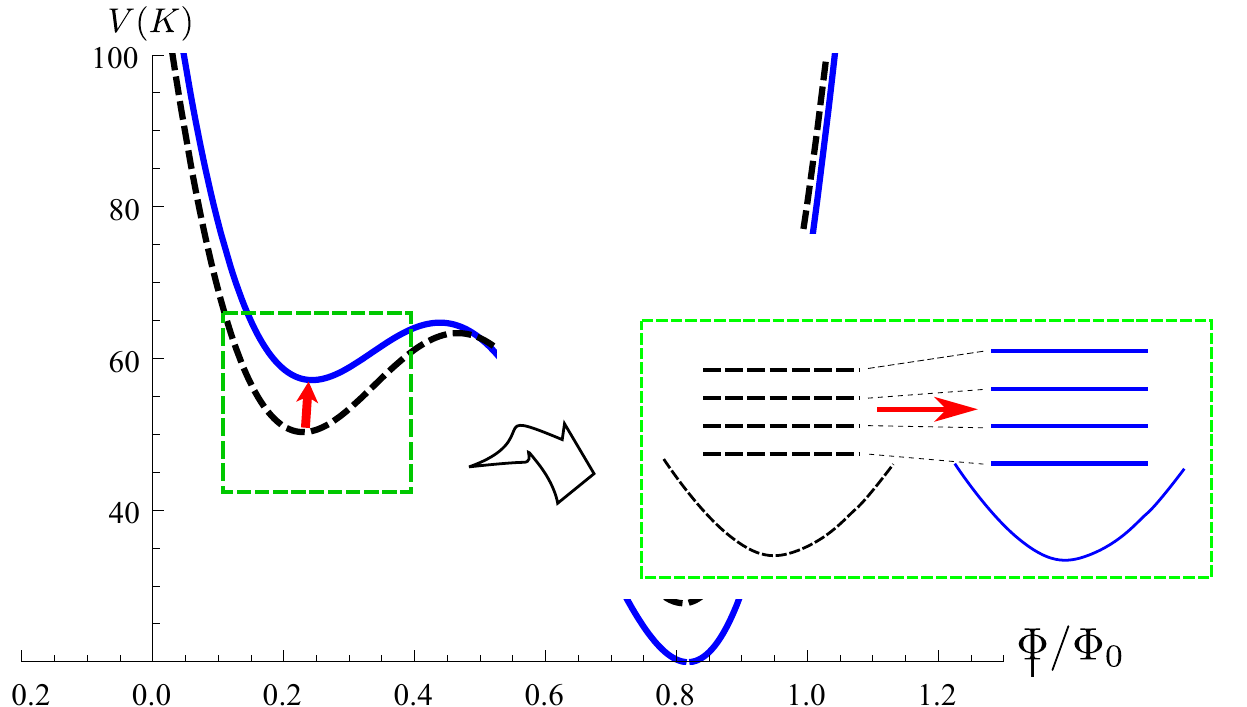}
\par\end{centering}

\protect\caption{Qualitative signatures of the topological phase in the potential energy
for the flux through the device at a given arbitrary bias. Dashed curves represent the trivial
limit when $\eta=0$, and blue curves a generic saturated (negative)
value of $\eta$ in the topological regime. (top) Modification of
the tunneling barrier induces a change in tunneling rates. (bottom)
Frequency shifts due to changes in curvature lead to shifts in the
discrete energy levels; \label{fig:Signatures}}
\end{figure}

At this point we are ready to discuss qualitatively the consequences
of this proposal. We look mainly at two possible signatures, namely,
changes in the tunneling rates and shifts in the harmonic oscillator
levels. The former might be probed in actual tunneling experiments
while the latter may be studied in spectroscopy or coherent tunneling
experiments. Fig.\ref{fig:Signatures} illustrates the two phenomena
and summarizes our main ideas. 

One sees how the competition between arbitrary integer and even integer
flux takes place. The $\left|\Phi\approx0\right\rangle $ well becomes shallower
while the $\left|\Phi\approx1\right\rangle $ one is deepened in comparison
with the trivial situation (this actually depends on the sign of $\eta$,
whose subtleties will be discussed further ahead, and, for now, we
keep in mind that the opposite sign would only bring an opposite scenario).

\begin{figure*}[t]
\begin{centering}
\subfloat[\label{fig:Casea}]{\begin{centering}
\includegraphics[scale=0.6]{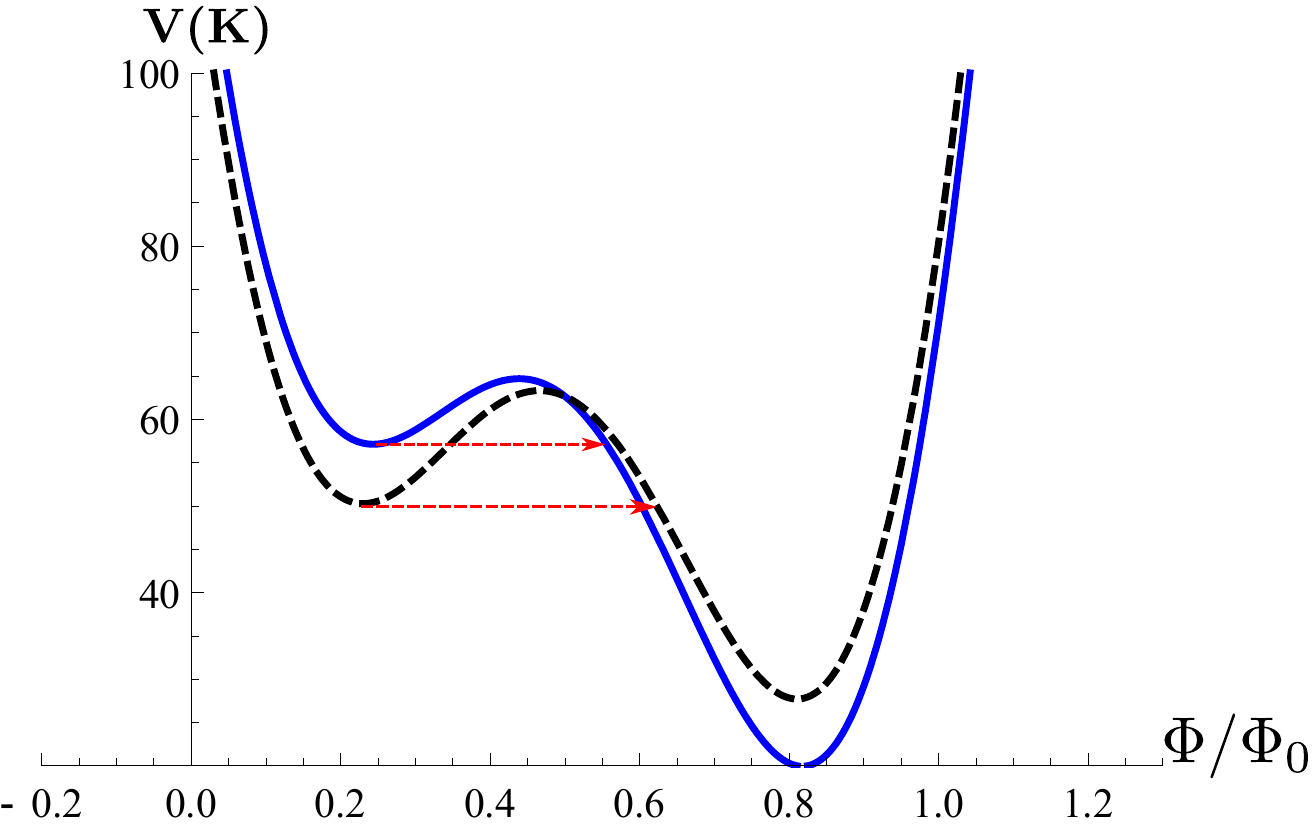}
\par\end{centering}

}\subfloat[\label{fig:Caseb}]{\begin{centering}
\includegraphics[scale=0.6]{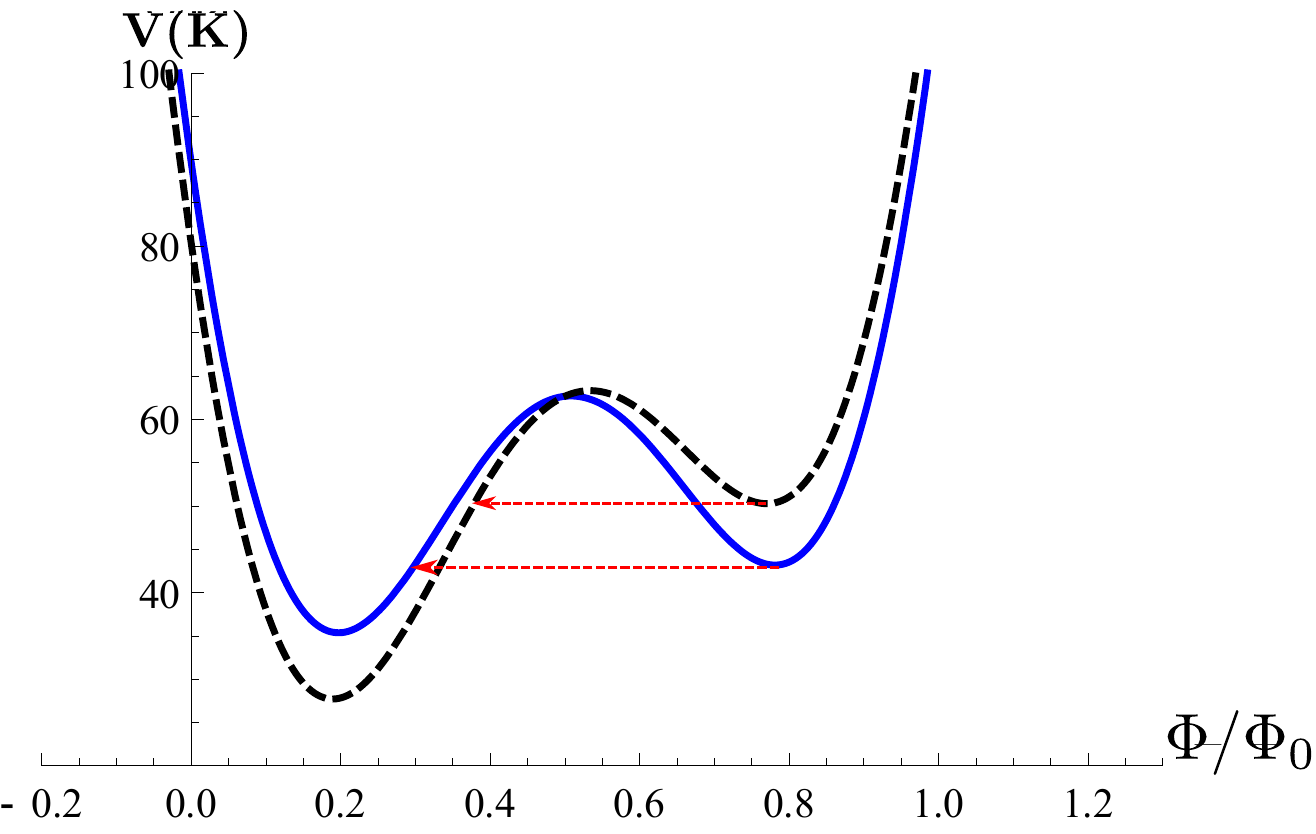}
\par\end{centering}

}
\par\end{centering}

\begin{centering}
\subfloat[\label{fig:Casec}]{\begin{centering}
\includegraphics[scale=0.6]{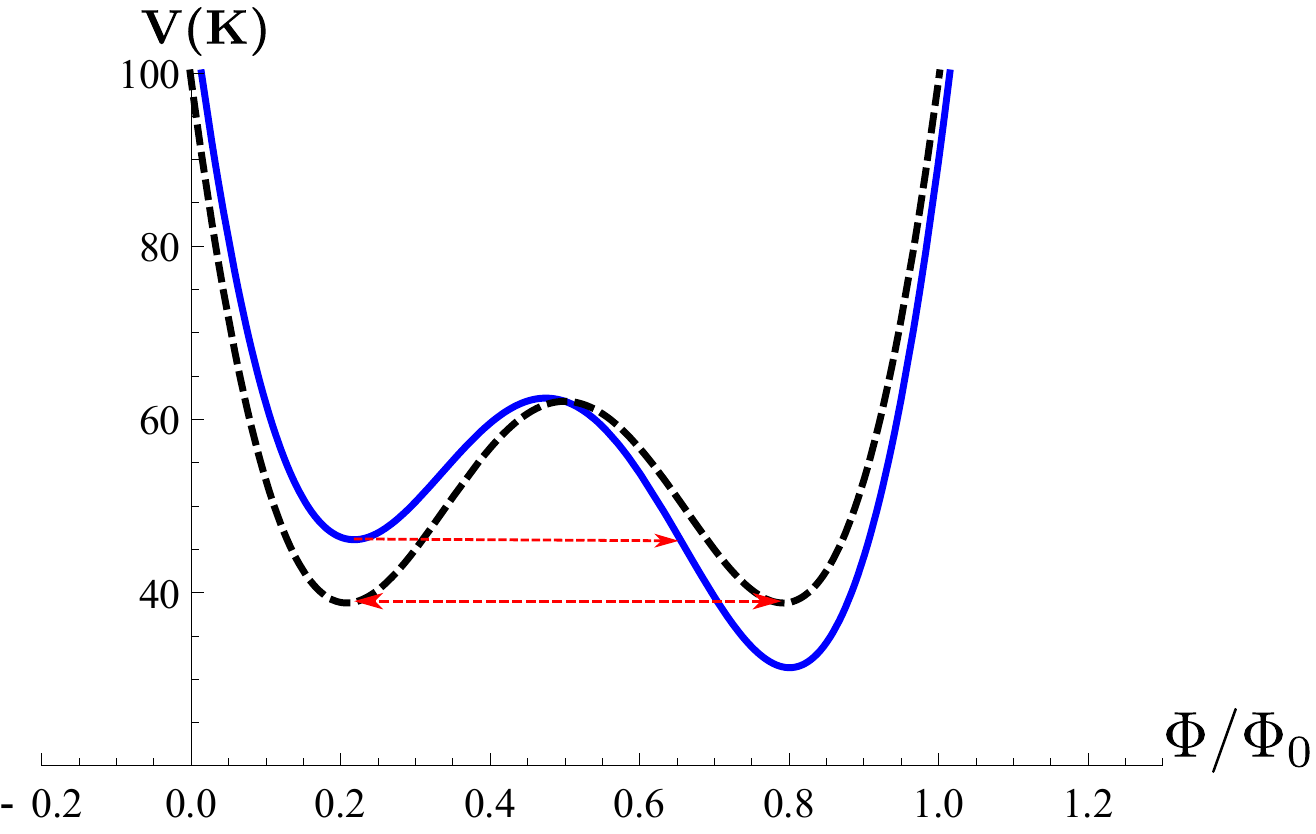}
\par\end{centering}

}\subfloat[\label{fig:Cased}]{\begin{centering}
\includegraphics[scale=0.6]{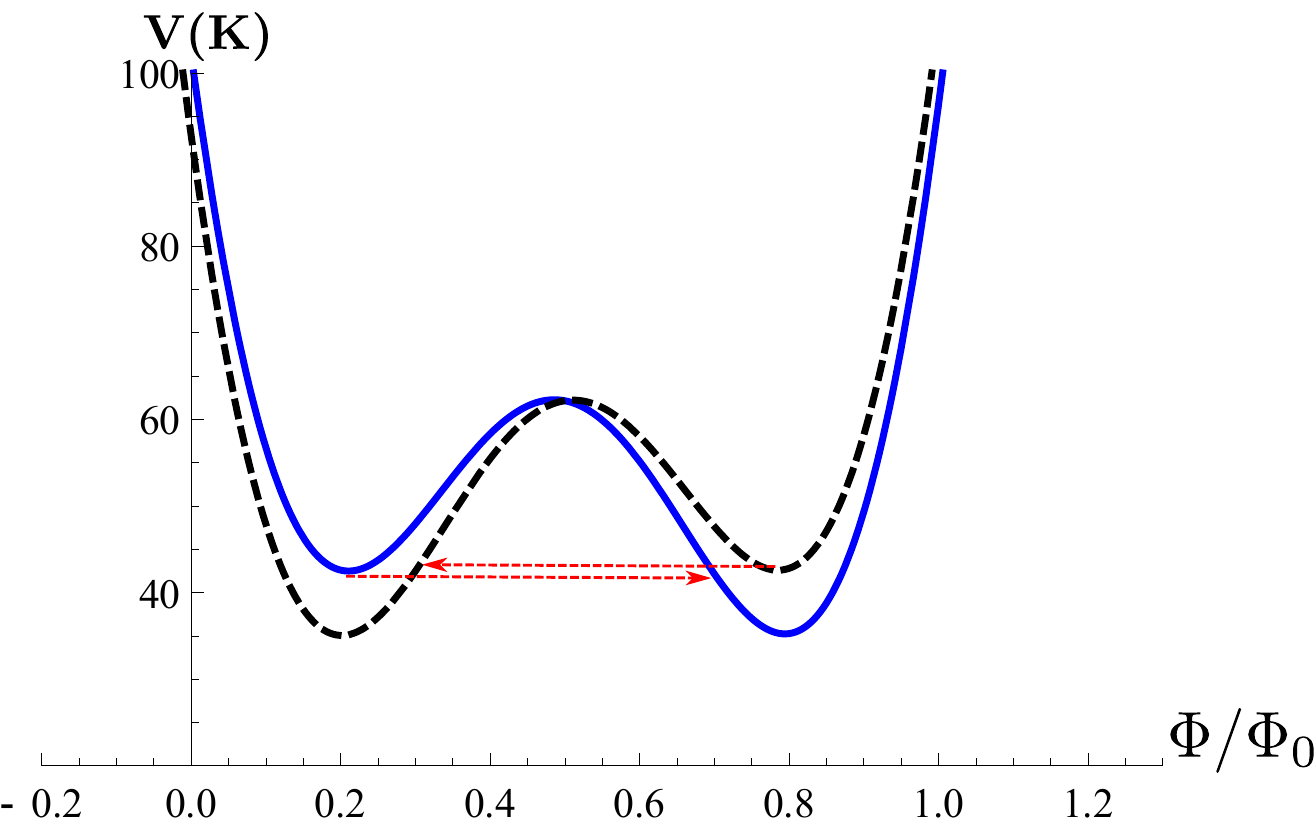}
\par\end{centering}

}
\par\end{centering}

\protect\caption{Different flux potentials as function of the external flux $\Phi_{X}$.
Black dashed curves represent $\eta=0$, blue solid curves represent
$\eta=-0.15$. $\beta_{L}=1.9$ in all cases. (a) ($\Phi_{X}=0.53$)
and (b) ($\Phi_{X}=0.47$) show the differences of tuning the external
flux around the symmetric (for the trivial regime) point $\Phi_{X}=0.5$.
In the case of figure (a) the tunneling rate is enhanced and in (b)
it is suppressed by the topological contribution. In case (c) $\Phi_{X}=0.5$
and we see how a ``cat state'' may be destroyed by lowering the
chemical potential into the topological regime. (d) shows how to build
a pump between the flux states by tuning the chemical potential into
and out of the topological regime.\label{fig:Cases}}
\end{figure*}

We may exploit many different schemes to study the consequences of
the topological regime. Fig.\ref{fig:Cases} gives some possibilities.
In all cases we take a physical value of $\beta_{L}=1.9$ and shift
$\Phi_{X}$ around the symmetric value for the non-topological regime
$\Phi_{X}=0.5$. For the sake of clearly describing the different
situations we take a value of $\eta=-0.15$. We note, however, that
its actual physical value might be much smaller.

In Figs.\ref{fig:Casea} and \ref{fig:Caseb} we see how situations
of enhancing and suppressing the tunneling in the well may be exchanged,
just by tilting the potential monitoring $\Phi_{X}$ from $0.47$
to $0.53$. These plots also make clear that the suppression or enhancement
are actually not symmetric around $\Phi_{X}=0.5$.

The last two cases of \ref{fig:Casec} and \ref{fig:Cased} present
very interesting possible applications. In Fig.\ref{fig:Casec}, we
see that starting with a symmetric potential, in a Schr\"{o}dinger's cat
state, we may transform the qubit into a simple classical bit or tune
the quantum state into a preferred value of the flux, just as a function
of the chemical potential. In Fig.\ref{fig:Cased} we see how to create
an adiabatic pump from the unit flux to zero flux and back by lowering
the chemical potential into the topological regime and raising it
back to the trivial situation.

\section{Quantitative results \label{sec:signatures}}

Now we discuss quantitatively the consequences of our proposals. In
order to keep ourselves on safe physical grounds, we use real parameters
and units taken from \cite{Friedman2000}. Namely, we have $C=1.04\times10^{-13}$ F,
$L=2.4\times10^{-10}$ H and keep in mind that experiments are done
at temperatures of the order of $10^{-2}$ K. We consider shallow potentials, as tunneling signatures
are our main goals. For comparison, we take the same values for the flux bias and $\beta_{L}$  parameters
when calculating frequencies and tunneling rates. Note however that, if resonance experiments are in mind,
 deeper wells should be preferred, as many oscillation levels may develop in this case. With all these in mind,
we may calculate the frequency and tunneling rate shifts as functions
of $\eta$. 

\begin{figure}[t]
\begin{centering}
\includegraphics[scale=0.55]{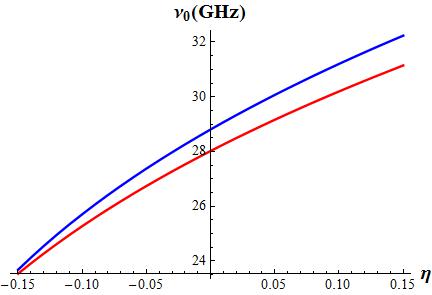}
\par\end{centering}

\protect\caption{Quantitative change in the oscillator frequencies $\nu_{0}=\omega_{0}/2\pi$ as a function of
the maximum value of $\eta$. We  parameters for
the SQUID as described in the text and $\Phi_{X}=0.53$,
$\beta_{L}=1.66$ (blue) and $\Phi_{X}=0.51$, $\beta_{L}=1.5$ (red). Comparing,  smaller
$\beta_{L}$ gives more expressive shifts in the frequency.\label{fig:FreqEta}}

\end{figure}

For the frequency we locate numerically the minima of the potential
as a function of $\eta$ and evaluate equation (\ref{eq:Freq}). In
Fig.\ref{fig:FreqEta} we plot the corresponding results, for both
positive and negative values of $\eta$ and for $\beta_{L}=1.5$ (in
red) and $\beta_{L}=1.66$ (in blue).

The results show that if we have a ratio of about 5\% between the
Majorana critical current and the parent SQUID critical current, we
may achieve a frequency shift of $\nu_{0}\sim$1 GHz. Different values of
$\beta_{L}$ shift the curves and, for small $\eta$, the slopes have small deviations.
It should be noted, however, that smaller values of $\beta_{L}$ 
actually also enhance $\eta$, since this is a ratio of the critical currents.

For the tunneling rate the calculation is slightly more involved.
In the non-dissipative limit it may be simply calculated from Callan
and Coleman's instanton calculation scheme \cite{PhysRevD.16.1762}.
In summary, it resumes to a saddle point approximation in the path
integral approach, considering paths in an inverted potential starting
and ending at the equilibrium point. The result reduces to
\begin{equation}
\Gamma=K\sqrt{\frac{B}{2\pi\hbar}}e^{-B/\hbar}\left[1+\mathcal{O}\left(\hbar\right)\right],\label{eq:rate}
\end{equation}
where
\begin{eqnarray}
B & = & \int_{-\infty}^{\infty}dt\left[\frac{C}{2}\dot{\Phi}_{Cl}^{2}+U\left(\Phi_{Cl}\right)\right]\\
 & = & 2\int_{0}^{\Phi_{w}}d\Phi\sqrt{2CU\left(\Phi\right)}
\end{eqnarray}
and
\begin{equation}
K=\sqrt{\frac{\det\left[-\partial_{t}^{2}+\omega_{0}^{2}\right]}{\det^{'}\left[-\partial_{t}^{2}+\omega_{0}^{2}+U^{"}\left(\Phi\right)\right]}}.
\end{equation}
Here $\Phi_{Cl}$ is the classical ``bounce'' solution in the saddle
point approximation and we used the equations of motion and energy
conservation to write $B$ independently from the exact solution $\Phi_{Cl}$.
The prime in the denominator determinant in $K$ means that the zero
eigenvalue should be omitted. The parameter $\Phi_{w}$ is the width of the
barrier the particle has to tunnel through and $\omega_{0}$ is again
the small oscillations frequency around the metastable minimum.

Instead of evaluating all the factors from the complicated potential
(\ref{eq:pot}), we follow the standard procedure and approximate it 
by a ``quadratic-plus-cubic'' potential,
\begin{equation}
U_{eff}\left(\Phi\right)=\frac{1}{2}C\omega_{0}^{2}\left[\Phi^{2}-\frac{\Phi^{3}}{\phi_{w}}\right].
\end{equation}
This is a very reasonable approximation \cite{Chang1984} and allows
us to write the tunneling rate in terms of dimensionless integrals
as
\begin{eqnarray}
B & = & 2C\omega_{0}\Phi_{w}^{2}\int_{0}^{1}dz\sqrt{\left[z^{2}-z^{3}\right]}.
\end{eqnarray}
The $K$ factor is dimensionless by definition and in the non-dissipative
limit that we are considering is given by $\sqrt{60}\sim7.75$ \cite{Chang1984,1967AnPhy..41..108L}.

\begin{figure}
\begin{centering}
\includegraphics[scale=0.55]{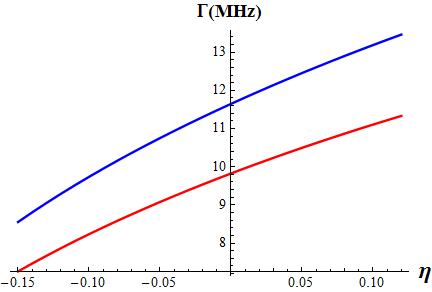}
\par\end{centering}

\protect\caption{Quantitative change in the tunneling rates as a function of the maximum
value of $\eta$. Thel parameters for the SQUID are, as described in the text, $\Phi_{X}=0.53$, $\beta_{L}=1.66$
(blue) and $\Phi_{X}=0.51$,  $\beta_{L}=1.5$ (red).\label{fig:TunRate}}

\end{figure}

Considering these, we may calculate numerically the minima and maxima
from the original potential, from which we can extract $\phi_{w}$.
The results are shown in Fig.\ref{fig:TunRate}. Again, for different
values of $\beta_{L}$, the slopes of the curves change. In particular, again
for $\eta=0.05$ (5\% ratio between critical currents), the tunneling
rate presents variations of $ \sim7\times10^{7}$ Hz.

Figures \ref{fig:FreqEta} and \ref{fig:TunRate} are our main
results. As mentioned before, some care must be taken regarding the depth of the wells, when
thinking of tunneling or resonance signatures. Frequencies must always be kept at 
small values, such that we avoid exciting undesirable quasi-particles in resonance experiments.  
Also, in tunneling experiments small tunneling rates demand too large coherence times
to observe the tunneling. It is remarkable
that, nevertheless, the expected signatures manifest themselves in such a way that one can
hope to actually measure them.

\section{Parity fluctuations and robustness \label{sec:Parityfluctuations}}

Now we turn to the problem of parity fluctuations. As discussed beforehand, in
this section we adopt a more heuristic point of view, discussing the
consequences more qualitatively and leaving a proper treatment of the
issue for future work. We will show that not only MQT measurements
are robust against parity fluctuations, but also that these fluctuations
enhance the differences in the tunneling rates.

The issue of fluctuations actually enriches the problem. We must now
be careful with the different time scales present, as discussed in
Ref.\cite{Fu2009}. The first time scale we need to think about is the
one which controls the parity fluctuation. The second time scale is
the one related to the evolution of the phase difference across the
JJ.

Two main sources of fermions are responsible for fluctuations, namely,
thermal excitation of quasiparticles or hopping from localized states in the bulk. 
These mechanisms are exponentially suppressed at low temperatures.
In the case when this process happens much faster than the evolution
of the SC phase (or the magnetic flux in our case), one has to be
careful and consider a proper thermal average of the current \cite{Kwon}.
A rigorous way to take this situation into account is to model phenomenologically
the parity fluctuation through a Fokker-Planck equation and consider
all possible combinations of tunneling processes \cite{Shu-Ping:1403.2747v1}.

We argue that this situation is improbable as follows. As long as
the phase is away from $\pi$, as in the two potential wells of our
potential, the different parity branches are far away from each other.
The gap separating them is large (of the order of the induced p-wave
gap magnitude) and the low temperatures, necessary for the macroscopic
quantum behavior of the device to manifest itself, should be enough
to suppress fluctuations \cite{Shu-Ping:1403.2747v1}. In an adiabatic
evolution of the phase through $\pi$, however, the system may access
the crossing point and even the lowest temperatures may introduce
corrections. 

It turns out that quantum tunneling do not describe adiabatic evolutions
of the phase. In a thermally activated phase slip, one might describe
the time within which the process take place by dividing the total
``distance'' traveled by the ``particle'' by its speed. In our
case, the situation is mode subtle. Tunneling rates describe the lifetime
of a metastable state but the transition itself is much faster. Since the
phase/flux behaves quantum mechanically as a ``position operator'',
defining how long the system spends at a given transition is not necessarily
straightforward. The problem of the tunneling time is controversial
and has been explored extensively \cite{RevModPhys.66.217}. A characteristic
time for this process is given by 
\begin{equation}
\tau=\frac{\phi_{w}C}{\hbar\kappa},
\end{equation}
where $\kappa$ is the imaginary momentum under the barrier,
\begin{equation}
\kappa=\sqrt{\frac{2C}{\hbar^{2}}V_{0}},
\end{equation}
and $V_{0}$ is the height of the barrier. In the mechanical picture,
this is the same as $md/(\hbar\kappa)$, where $m$ the mass of the
particle and $d$ the width of the potential barrier. In our case,
\begin{eqnarray}
\kappa & \sim & \frac{C\omega_{0}\phi_{w}}{\hbar},\\
\Rightarrow\tau & \sim & \frac{1}{\omega_{0}}\sim10^{-7}s,
\end{eqnarray}
but it has been argued that it is not always that this time has physical
significance \cite{RevModPhys.66.217}.

\begin{figure}
\subfloat[\label{fig:ZMspectrum}]{\includegraphics[scale=0.52]{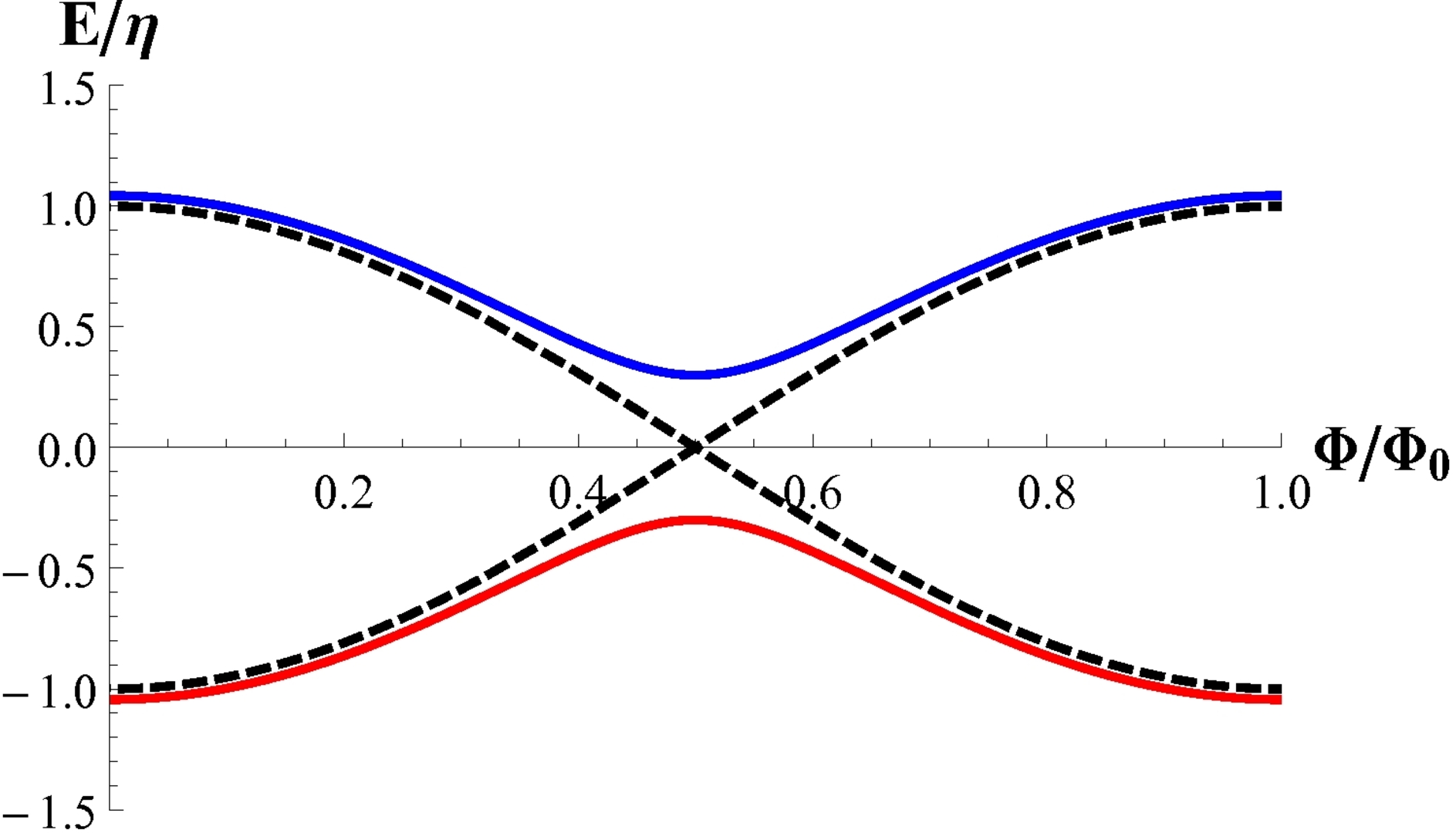}

}

\subfloat[\label{fig:ParityBrokenFluxPot}]{\includegraphics[scale=0.58]{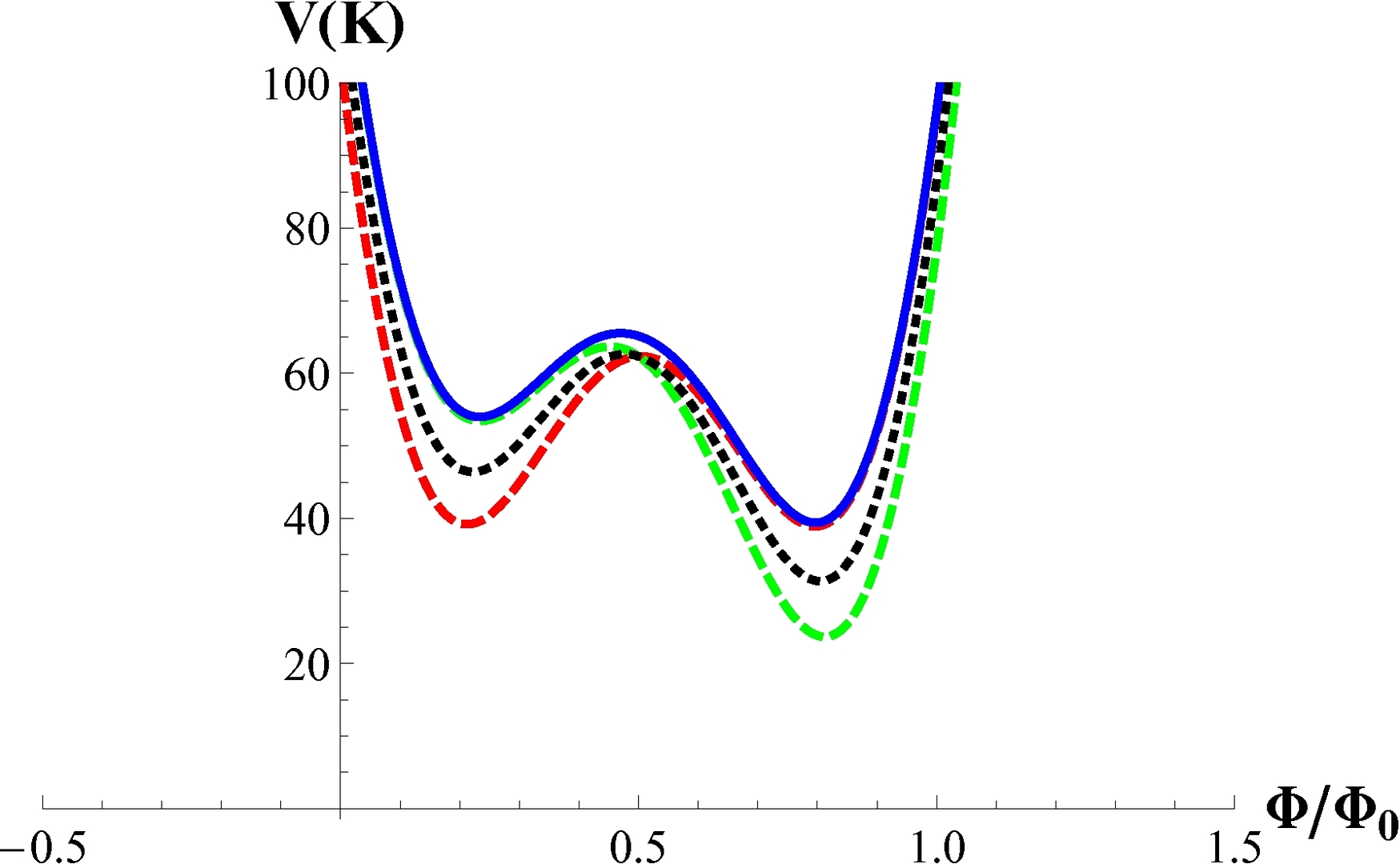}

}

\protect\caption{(a) Andreev spectrum for an open wire in the presence of fermionic
parity breaking terms. The black dashed curves represent the parity
preserving limit; (b) Comparison between the trivial ($\eta=0$,
black, dotted), topological and parity conserving ($\eta=-0.15$ red,dashed
and $\eta=0.15$ Green,dashed) and topological and parity breaking
($\left|\eta\right|=0.15$ blue) potential energies. We use $\beta_{L}=1.9$
and $\delta=0.3$. Notice how the blue curve starts overlapping
with the green one and then exchanges at $\Phi=0.5$ to overlapping
with the red one. The introduction of a parity breaking contribution
violates the $4\pi$ periodic signature and mixes the potential profiles
at  $\Phi=0.5$. Tunneling may be enhanced
or suppressed due to the exchange between parity branches.}
\end{figure}

A safe claim is that, for sure, the time evolution of the phase across
the junction is not adiabatic in a tunneling experiment and we focus
our attention now onto the picture that parity fluctuations are slower
than the tunneling process. In this case, the parity symmetry breaking
will allow for the coupling and gapping of the two parity eigenstates.
The simple way to model this is to consider the low energy projected
Hamiltonian as 
\begin{eqnarray}
H\left(\Phi\right) & = & 2\eta\left(2i\gamma_{1}\gamma_{2}\cos\frac{\Phi}{2}+\delta_{1}\gamma_{1}+\delta_{2}\gamma_{2}\right)\nonumber \\
 & = & \eta\left(\left(2c^{\dagger}c-1\right)\cos\frac{\Phi}{2}+\left[\left(\delta_{1}+i\delta_{2}\right)c+H.c.\right]\right),\nonumber \\
\end{eqnarray}
where the $\gamma_{1,2}$ are the Majorana modes at the wire's ends,
defining a two-level system given by the complex fermions
\begin{equation}
\gamma_{1}=\frac{c+c^{\dagger}}{2},\gamma_{2}=\frac{c-c^{\dagger}}{2i}.
\end{equation}
This Hamiltonian is to be regarded as a mean-level description of
the Hamiltonian in \cite{Fu2009}, which describes inelastic processes
responsible for parity flipping events. We just want to study the
qualitative features of this system and, as such, capture them into
the parameters $\delta_{12}$, which are normalized by $\eta$. Clearly,
for $\delta_{1,2}\neq0$, this Hamiltonian does not
conserve the fermionic parity and a gap opens up in the Andreev bound
states spectrum as illustrated in Fig.\ref{fig:ZMspectrum}.

For non-vanishing $\delta$, a gap is opened at the two different
parity states, restoring the $2\pi$ periodicity to the energy levels.
The new potential for the SQUID is then given by
\begin{eqnarray}
U\left(\phi\right) & = & U_{0}\left[\frac{\left(2\pi\left(\Phi-\Phi_{X}\right)\right)^{2}}{2}\right.\label{eq:ParityBrokenPot}\\
 &  & \left.-\beta_{L}\left(\cos2\pi\Phi+\left|\eta\right|\sqrt{\left|\delta\right|^{2}+\cos^{2}\pi\Phi}\right)\right].\nonumber 
\end{eqnarray}
In this case, a gap of order $\delta$ opens up in the Andreev states
spectrum. This situation actually leads to two consequences. Firstly,
$\delta$ lifts the degeneracy of the trivial potential at $\Phi=0.5$.
This slight raising or lowering of the potential barrier has little
consequences for tunneling rates. This can be understood by noticing
that the coefficient $B$ depends linearly on the frequency $\omega_{0}$
and quadratically on $\phi_{w}$. The raising/lowering of the barrier
height by $\delta$ mainly affects the frequency, leaving the width
intact and generating very small corrections to the tunneling rate.

There is, however, a second important point. Figures \ref{fig:ZMspectrum}
and \ref{fig:ParityBrokenFluxPot} summarizes the situation. The spectrum
has no $4\pi$ periodicity anymore. One sees, in a closer analysis,
that effectively the new potential interpolates between the two parity
states, exchanging to opposite parities as $\Phi$ crosses $0.5$.
Consider for definiteness that the system is prepared in the positive
$\eta$ state. The tunneling barrier width $\phi_{w}$ now has become
wider and this clearly gives a substantial deviation to the tunneling
rate. The opposite would happen if one started from the red curve,
with a shrinking of $\phi_{w}$ but still a substantial deviation
to the tunneling rate from the trivial case would take place. In many
realizations one would have to average over the two possibilities.
\begin{figure}[t]
\includegraphics[scale=0.62]{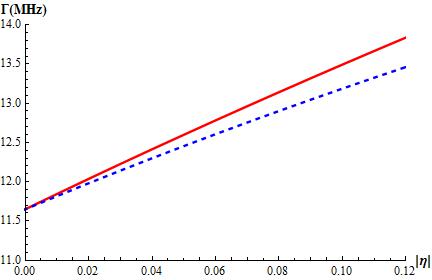}

\protect\caption{Violation of parity conservation enhances the effects of the topological
phase in the tunneling rate signatures. Here $\beta_{L}=1.66$,
$\eta=-0.15$ and we compare $\delta=0.3$ (red) and $\delta=0$ (blue,
dashed.)}
\end{figure}

Clearly, at $\eta=0.05$, if $\delta=0.3$ we have an enhancement of the tunneling rate
as compared to $\delta=0$. This shows
that a slow parity fluctuation acts in favor of the tunneling signatures
for detecting TSC in rf-SQUIDs. One notices, on the other hand, that
the flux pumps described in Fig.\ref{fig:Cased} and the ``transmutation''
of a qubit into a classical bit from \ref{fig:Casec} are not possible
in this situation.

\section{$\pi$-junction\label{sec:pi-junction}}

We now briefly extend the discussion of the previous sections to the
case of $\pi$- junctions. We have been considering the total phase
across the junction to be totally controlled by the flux through the
loop. Now we assume that the SC also builds a $\pi$ phase across
the junction. 

The idea follows from the discussion in\cite{Lucignano:1302.5242v2},
where the Andreev spectrum of the Kitaev wire is studied thoroughly.
This discussion encompasses several cases, including open and closed
wires and finite size effects. It is seen that both open and closed
limits may be made symmetric under proper conditions, such that the
potential becomes sinusoidal.

\begin{figure}[t]
\includegraphics[scale=0.60]{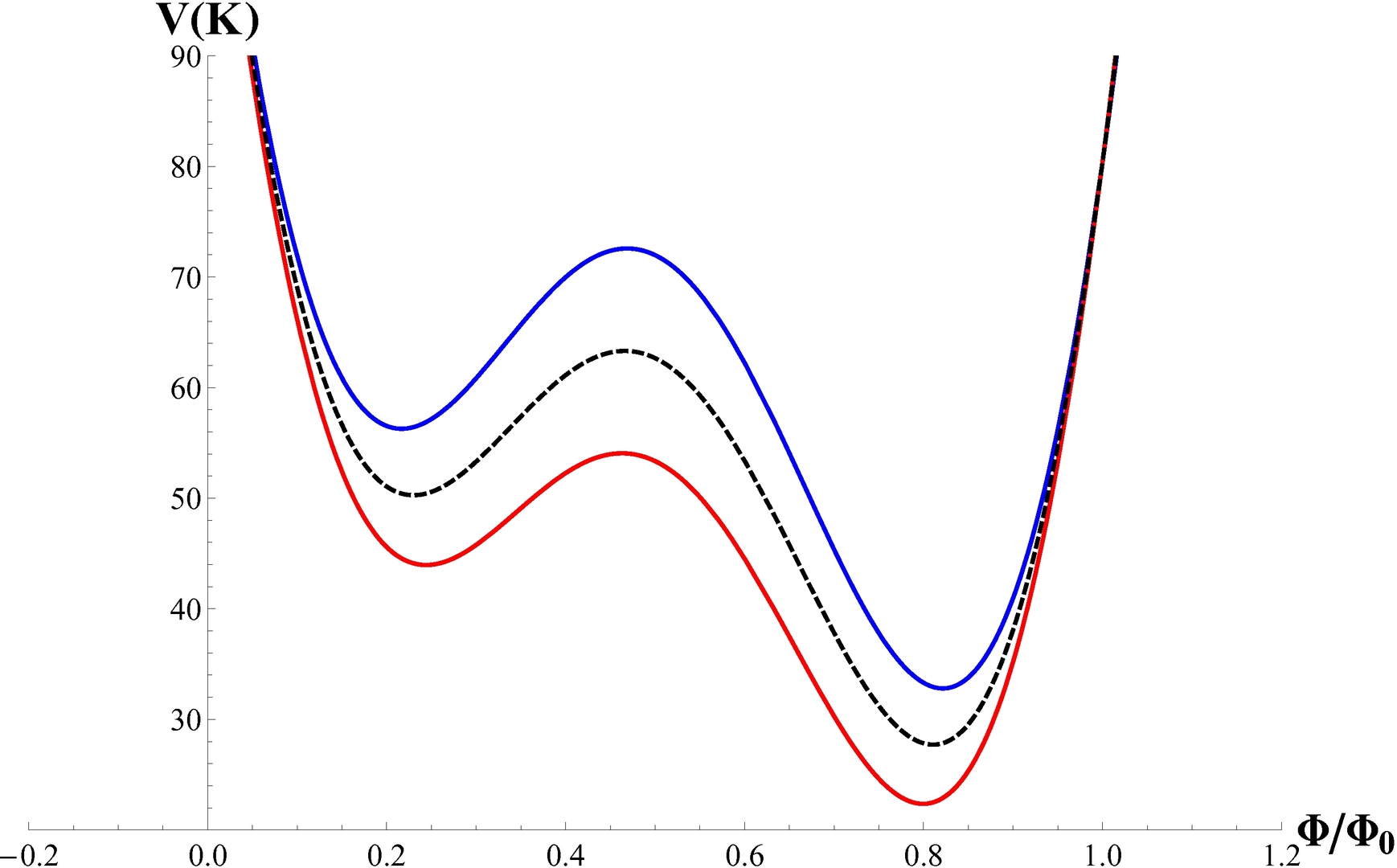}

\protect\caption{Potential energy for the $\pi$-junction wire in the symmetric limit.
$\beta_{L}=1.9$, and $\Phi_{X}=0.53$. Here $\eta=-0.15$ for the
blue curve and $\eta=0.15$ for the red curve.\label{fig:SinePot}}

\end{figure}

The flux dynamics is then described by the following potential,
\begin{eqnarray}
U\left(\phi\right) & = & U_{0}\left[\frac{\left(2\pi\left(\Phi-\Phi_{X}\right)\right)^{2}}{2}\right.\\
 &  & \left.-\beta_{L}\left(\cos2\pi\Phi+\eta\left(\mu\right)\sin\pi\Phi\right)\right].\nonumber 
\end{eqnarray}
Competition between even and odd minima now is absent. Both wells
are lifted or lowered, depending on the parity eigenstate, as illustrated
in Fig.\ref{fig:SinePot}. This is however, different from the parity
broken case discussed in the last section, as the topological part
of the potential is clearly $4\pi$ periodic. Tunneling between different
branches in this situation seems to give smaller differences than
those in our original case.

\begin{figure}[t]
\begin{centering}
\includegraphics[scale=0.55]{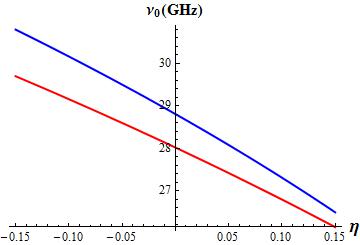}
\par\end{centering}

\begin{centering}
\includegraphics[scale=0.55]{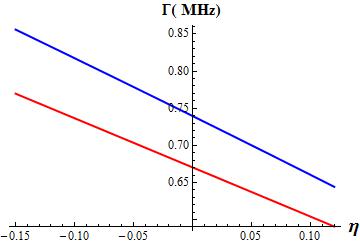}
\par\end{centering}

\protect\caption{Frequency and tunneling rates for the $\pi$-junction. Again $\Phi_{X}=0.53$,  $\beta_{L}=1.66$
(blue) and $\Phi_{X}=0.51$, $\beta_{L}=1.5$ (red)\label{fig:SineShifts}. The slopes
are inverted in comparison with \ref{fig:FreqEta} and \ref{fig:TunRate}
but again the topological phase again leaves its signatures. The $\pi$-junction
is less susceptible to the changes in the absolute value of $\beta_{L}$.}

\end{figure}

Similarly to the parity broken situation, this case is not so interesting
as the open wire in the sense that we cannot engineer flux pumps and
qubits that may be tuned into simple bits as in the discussion of
Fig.\ref{fig:Cases}. The typical behavior of the signatures in frequency
and tunneling rates are quite remarkable  in this case, since they
are the opposite from the case of the open wire. As one can see in
Fig.\ref{fig:SineShifts}, the roles of positive and negative parity
are now inverted. 

One might use this in the parity conserving limit to distinguish between the 
physics of closed or open wire. When parity conservation is broken, the differences 
between the two cases may disappear.

\section{Conclusions \label{sec:Conclusions}}

We have studied topological rf-SQUIDs and analyzed the consequences
of the topological phase in MQT experiments. The resonance phenomena
proposed are expected to give valuable evidence of the existence of
the superconducting topological phase. These proposals have the additional
advantage of  avoiding some of the difficulties that arise in the interpretation 
of transport experiments. 

From a phenomenological model, we show that the crossover between trivial and topological 
phases may be probed by looking at shifts in the tunneling rates and oscillation frequencies of SQUIDs in the macroscopic
quantum regime.  The limit of small critical currents ratio $\eta$ is analyzed, although
one may hope to  have some control over its magnitude. The introduction of a second
loop in the parent SC, for example, allows for tuning $\beta_{L}$
\cite{Friedman2000} and, as such, $\eta$. Moreover, the topological
part of the critical current itself depends on the junction length,
as well as on the strength of the pairing \cite{Alicea2012,Fu2009}.

Parity breaking is found to work in favor of the detection of the
topological phase, as long as the fluctuations are slower than the
phase evolution. As we argued, due to the non-adiabatic nature of
the flux tunneling process, it is reasonable to expect this to be
the general case. This is in accordance with the quantitative results
in the limit of fast phase slipping in the context of SC-QSHI-SC biased
junctions from\cite{Shu-Ping:1403.2747v1}. In the latter, the authors
also propose to detect the TSC phase in a SC-QSHI-SC junction by addressing
the consequences of strong parity fluctuations in thermally activated
phase slipping. The authors claim, on the other hand, that in the
zero temperature limit, telegraph noise (from parity fluctuation events)
averages the voltage across the device to zero at low external currents.
Quantum tunneling implies that a drift will actually remain, even
in the very low temperature limit, and a finite voltage should develop
even for very small bias currents.

The setup in which we described our ideas requires fine tuning of the device to achieve 
topological superconductivity, as well as means for tuning into and out of
the topological phase. For simplicity, we restricted ourselves to a toy model based on  
the SC coupled to strong SO wire device of \cite{PhysRevLett.105.077001,PhysRevLett.105.177002}. 
Our ideas, however, do not rely much on the geometry of the device. Corbino geometries like 
described in Fu and Kane \cite{Fu2009} or, as discussed, open SC-QSHI-SC current biased 
junctions are most likely the proper systems in which this physics should be studied. These
junctions are very promising systems for the realization of the TSC
phase, among other reasons, for avoiding the necessity of chemical
potential fine tuning.

 Tuning between phases in these situations is also easily achieved by the application of
in-plane magnetic fields. These devices are, however, characterized by larger Majorana contributions
to the critical currents as compared to the $2\pi$ periodic one, that is, these
systems work in the regime of large $\eta$. In this case, MQT between
even integer flux states may be achieved, with small shifts from the expected rates due to the
non-topological currents. Parity fluctuations will likely spoil the $4\pi$ periodicity of the potential again, restoring the 
tunneling barrier width to a much smaller value corresponding to non-topological regime.

Besides the above discussed points, we considered the $\pi$-junction limit which might be of relevance
under the light of the new proposals of development of TSC from d-wave
parent SCs \cite{PhysRevB.86.235429,PhysRevB.87.014504}.

\section*{Acknowledgments}

The authors acknowledge fruitful discussions with S. Ryu, T. Hughes
and S.-P. Lee. The authors are particularly indebted with and grateful
for enlightening discussions with J. C. Y. Teo. This work was supported
by FAPESP under grants 2009/18336-0 and 2012/03210-3. VS
acknowledges financial support from DOE DE-FG02-07ER46453.  AOC is supported
by Instituto Nacional de Ci\^encia e Tecnologia  de  Informa\c{c}\~ao Qu\^antica
(INCT-IQ) under grants CNPq - 610020/2009-9 and FAPESP - 2008/57856-6.

\bibliographystyle{apsrev4-1}

%

\end{document}